\pdfoutput=1
\documentclass[reprint,linenumbers,aip, graphicx,amsmath]{revtex4-1}
\usepackage{graphicx}
\usepackage{dcolumn}
\usepackage{bm}
\usepackage{bm}
\usepackage{hyperref}
\hypersetup{
   colorlinks,=true,
    citecolor=blue,
    filecolor=blue,
    linkcolor= blue,
   urlcolor=blue
}
\begin{document}


\title{Charged basal stacking fault (BSF) scattering in nitride semiconductors} 



\author{Aniruddha Konar}
\email[]{akonar@nd.edu}
\affiliation{ Department of Physics, University of Notre Dame, Notre Dame, USA 46556}
\affiliation{ Midwest Institue of Nanoelectronics Discovery (MIND), Notre Dame, USA 46556}
\author{Tian Fang}
\affiliation{  Department of Electrical Engineering, University of Notre Dame, Notre Dame, USA 46556}
\affiliation{ Midwest Institue of Nanoelectronics Discovery (MIND), Notre Dame, USA 46556}
\author{Nan Sun}
\affiliation{ Department of Physics, University of Notre Dame, Notre Dame, USA 46556}
\author{ Debdeep Jena}
\affiliation{ Department of Electrical Engineering, University of Notre Dame, Notre Dame, USA 46556}
\affiliation{ Midwest Institue of Nanoelectronics Discovery (MIND), Notre Dame, USA 46556}


\date{\today}

\begin{abstract}
 A theory of charge transport in semiconductors in the presence of basal stacking faults is developed. It is shown that the presence of basal stacking faults leads to anisotropy in carrier transport. The theory is applied to carrier transport in non-polar GaN films consisting of a large number BSFs, and the result is compared with experimental data.
\end{abstract}

\maketitle 
The III-V nitride semiconductors and related compounds have attracted immense attention for  optoelectronic devices \cite{WaltereitNature00, BookPiprek} covering a wide range of the electromagnetic spectrum as well as high-speed, high temperature electronic devices. For c-plane grown nitrides, built-in polarization field, although advantageous for two dimensional electron gas (2DEG) formation in transistors \cite{BookJena}, reduces the overlap of electron-hole wavefunction,  reducing the oscillator strength for radiative transitions. Moreover, c-plane based enhancement mode transistors suffer from low threshold voltages \cite{CaiEDL05} ($V_{th} \sim 1$ V) impeding their applications in safe circuit operation. A potential way to eliminate these effects is to grow nitrides in non-polar directions. However, heteroepitaxial, non polar and semi-polar nitrides films grown on lattice mismatched substrates   contain $n_{SF}\sim$ 1-10 $\times 10^{5}$/cm basal stacking faults (BSFs) parallel to the [0001] direction \cite{BenjaminJEM05,HsiaoJAP10, UenoJJAP10}. Directionally dependent transport measurements in highly faulted non-polar nitride films show strong mobility anisotropy for both electron and hole transport, with mobility parallel to the [0001] direction significantly lower relative to the in-plane [11$\bar{2}$0]  mobility \cite{McLaurinJAP06,McLaurinRRL07,BaikIEEE10} . This anisotropy has been associated with scattering from BSFs qualitatively, although no quantitative model exists. In this letter, we develop a theory of carrier scattering phenomena from BSFs and a quantitative estimate of the transport anisotropy is presented with comparison to experimental data.\\
\begin{figure}[b]%
\includegraphics*[width=99 mm]{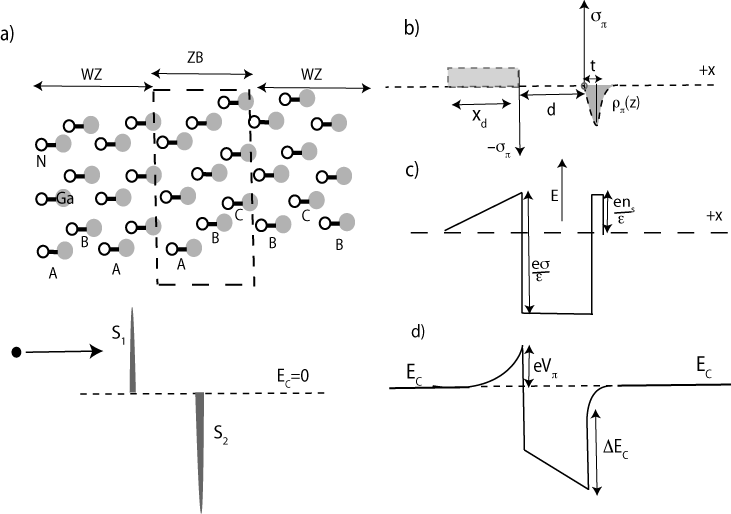}
\caption{%
  a) structure of basal stacking faults , b) charges  across the basal stacking fault, c) schematic diagram of electric field, d) conduction band diagram including band-offset, e) delta-function model of barrier and QW.}
\label{Fig1}
\end{figure}
\par
 Stacking faults (SF) are 2D structural defects associated with heteroepitaxial growth on lattice-mismatched substrates.  Most prevalent SFs in wurtzite (WZ)  nitrides is of the $I_{1}$-type requiring lowest energy of formation \cite{ StampflPRB98}. Scattering-contrast transmission microscope (TEM) imaging revealed \cite{BenjaminJEM05} that the SFs in GaN are primarily of the $I_{1}$-type which corresponds to a stacking sequence of (0001) basal plane $\cdots ABAB{\bf{ABC}}BCBC\cdots$. A BSF can be thought as a thin zincblende (ZB) layer (up to three monolayers thick) perfectly inserted in the WZ matrix without broken bonds as shown in Fig.\ref{Fig1}a). The built-in polarization difference between ZB and WZ structures will result in bound sheet charges $\pm\sigma_{\pi}$ at each interface of the BSF layer. Consequently, the band edge will bend near the BSF. Fig.\ref{Fig1}b) shows a typical conduction band diagram around a BSF for a n-doped GaN film along with the charges that are formed in the direction perpendicular to the BSF plane. The bending of conduction band edge inside the BSF is due the electric field resulting from the polarization bound charge, whereas, bands bend outside the BSF due to the accumulation and depletion of mobile charges. We approximate the accumulation charge as a sheet charge of density $n_{s}$ at the centroid ($t$) of the charge distribution, i.e. $\rho_{\pi}(x)=en_{s}\delta(x-x_{0})$, where $x_{0}=d+t$, $d$ being the width of a BSF and $e$ is the electron charge. If $x_{d}$ is the width of the depletion region, charge neutrality requires $n_{s}=x_{d}N_{d}$, where $N_{d}$ is the donor density. Energy conservation across the BSF leads to the relation
\begin{equation}
\frac{eN_{d}x_{d}^{2}}{2\epsilon_{s}}+\frac{en_{s}t}{2\epsilon_{s}}=\Big(\frac{e\sigma_{\pi}}{2\epsilon_{s}}-
\frac{en_{s}}{2\epsilon_{s}}\Big)d ,
\label{Eq1}
\end{equation}
where, $\epsilon_{s}$ is the permittivity of the semiconductor.  In the limit $x_{d}\gg (t,d)$,  the depletion length is $x_{d}\simeq \sqrt{\sigma_{\pi}d/N_{d}}$. The depletion width $x_{d}$ increases with decreasing donor density and there exists a critical donor density $N_{d}^{cr}$ for which,  $x_{d}$ becomes equal to the distance between two stacking faults. As a result the whole channel is depleted, and conduction ceases along the $x$ direction. The critical donor density above which onset of conduction along $x$ direction occurs  is given by $ N_{d}^{cr}\ge\sigma_{\pi} dn_{SF}^{2}$. For a typical fault density $n_{SF}=10^{5}$/cm and $d=0.8 $ nm for GaN \cite{RebanePSS97}, the critical doping density is $N_{d}^{cr}=10^{16}$/cm$^{3}$. In the rest of the paper, we will work in the regime where $N_{d}\gg N_{d}^{cr}$.\\
\par
In addition to the polarization charges at interfaces, the conduction band offset ($\Delta E_{c}$) between ZB and WZ structures leads to a quantum well (QW) in the fault region as shown in Fig.\ref{Fig1}b). For an applied bias, electrons tunnel through the barrier (in the depletion region) and QW (in the fault region)  and then diffuse in the space between two consecutive stacking faults. To model the transmission coefficient of tunneling, we approximate the barrier and the QW as two delta functions (as shown in Fig.\ref{Fig1}e) of strengths $S_{1}=eV_{\pi}x_{d}$ and $S_{2}=(\Delta E_{c}+eV_{\pi})d$; where $V_{\pi}=\sigma_{\pi}d/2\epsilon_{s}$. The energy dependent coefficient of transmission through a single delta function potential is an analytically solvable problem\cite{GriffithsQM}.  If $T_{tr,1}(\epsilon)$ and $T_{tr,2}(\epsilon)$ are the transmission coefficient from the barrier and the QW respectively, then the total  coefficient of transmission across the BSF is given by $T_{tr}(\epsilon)=T_{tr,1}(\epsilon)T_{tr,2}(\epsilon)$: 
\begin{equation}
T_{tr}(\epsilon)=\Big[\frac{1}{1+\frac{m^{\star}S_{1}^2}{2\hbar^{2}\epsilon}}\Big]\times\Big[\frac{1}{1+\frac{m^{\star}S_{2}^2}{2\hbar^{2}\epsilon}}\Big],
\label{Eq2}
\end{equation}
where, $\epsilon$ is the energy of the incoming electron and $m^{\star}$ is the effective mass at  the band edge. In SF-free structures, the elemental current component along the $x$ direction for an electron with velocity $v_{k}$ is given by $j_{x}^{k}=ev_{k}^{x}\delta f_{k}$; where $\delta f_{k}$ is the modification of equilibrium Fermi-Dirac distribution caused by an applied electric field $F_{appl}$. In the presence of SFs, a part (given by Eq. 2) of these carriers is transmitted through the barrier and the QW leading to an effective elemental current density $j_{k}^{eff}=T_{tr}(\epsilon)j_{k}^{x}$. Then, the total current density in the presence of SFs is given by
\begin{equation}
J_{x}=2e\int \frac{d^{3}k}{(2\pi)^3}T_{tr}(\epsilon)v_{k}^{x}\delta f_{k},
\label{Eq3}
\end{equation}
where, $v_{k}^{x}$ is the $x$ component of electron group velocity in the state $|{\bf{r,k}}\rangle$ and the factor 2 takes spin degeneracy into account. For a small applied electric field along the $x$ direction, under the relaxation time approximation (RTA), 
$\delta f_{k}\approx(-\partial f_{0}/\partial\epsilon)\tau(k)v_{k}^{x}F_{appl},$
where, $f_{0}$ is the equilibrium Fermi-Dirac distribution function and $\tau(k)$ is the momentum relaxation time due to scattering from impurities, phonons etc present in the material. Integrating Eq.\ref{Eq3} over all k-space and defining conductivity by the relation $J_{x}=\sigma(n,T) F_{appl}$, one obtains the carrier concentration ($n$) and temperature ($T$) dependent conductivity in the presence of BSFs.  The expression for conductivity across the BSF is then given by
\begin{equation}
\sigma_{xx}(T,n)=\frac{3ne^{2}}{8\pi}\frac{\int d\epsilon\epsilon^{1/2}\cosh^{-2}(\epsilon/2k_{B}T)T_{tr}(\epsilon)\tau(\epsilon)(v_{x}^{k})^2}{\int d\epsilon\epsilon^{3/2}\cosh^{-2}(\epsilon/2k_{B}T)},
\label{eq4}
\end{equation}
where $T$ is the equilibrium temperature, $k_{B}$ is Boltzmann constant.  Since the BSF does not break the periodic symmetry in the $y$ and $z$ directions, inserting $T_{tr}(\epsilon)=1$ for all energies,  a similar expression can be obtained for conductivity ($\sigma_{yy}$) along the $y$ direction. For hole transport, $\Delta E_{c}$  should be replaced by valence band offset ($\Delta E_{v}$) in the strength of delta function potential $S_{2}$,  and a hole effective mass should be used instead of the electron effective mass.\\
\par
Equation (\ref{eq4}) is the central finding  of this work as it allows us to calculate experimentally measurable quantities such as conductivity and mobility in the presence of BSFs. As an application of the formalism constructed above we investigate charge transport in p-type {\it{m}}-plane (1$\bar{1}$00) GaN in the presence of BSFs, and compare the results with previous reported experimental data. Inclusions of different scattering mechanisms are necessary to evaluate the energy dependent momentum relaxation time $\tau(\epsilon)$ appearing in Eq.\ref{eq4}.  Due to the high activation energy of acceptors ( Mg for p-type GaN has activation energy $\sim$ 174 meV \cite{McLaurinJAP06}), a large fraction of the dopants remain neutral even at room temperature, acting as neutral impurity scatterers to hole transport. The fact that a high doping density is required to achieve appreciable hole concentrations results in high density of neutral impurities (NI) making neutral impurity scattering dominant even at room temperature. Accounting for the zero-order phase shift \cite{ErginsoyPR50}, the momentum relaxation time of neutral impurity scattering is  $\tau_{NI}^{-1}= 20N\hbar a_{0}/m^{\star}$; where $N$ is the density of neutral impurities and $a_{0}=4\pi\epsilon_{s}\hbar^{2}/m^{\star}e^{2}$ is the effective Bohr radius. Momentum relaxation time  due to ionized impurity scattering ($\tau_{imp}$) is calculated following the method of Brooks-Herring (BH)\cite{BrooksPR51}. The fact that $\tau_{imp}\sim n_{imp}^{-1}$; where $n_{imp}$ is the density of ionized impurities, makes ionized impurity scattering important at high carrier concentrations (ionized dopant concentrations).  For electron-optical phonon momentum-relaxation time ($\tau_{ph}$), only phonon absorption has been considered due to the high optical phonon energy  ($ E_{op}= 0.092$ eV $\gg k_{B}T$) of GaN. The resultant momentum-relaxation time is calculated using Mathiessen's rule; i.e. $\tau(\epsilon)^{-1}=\tau_{NI}^{-1}+\tau_{imp}^{-1}+\tau_{ph}^{-1}$. For numerical calculations, we assume the valence band offset $\Delta E_{v}$=0.06 meV \cite{StampflPRB98}. The hole effective mass of GaN is not well known and a wide range $m_{h}^{\star}= 0.4-2.4 m_{0}$ ($m_{0}$ is the rest mass of a bare electron) is found in existing literature \cite{BookPiprek}. In this work, we have assumed $m_{h}^{\star}\sim1.8 m_{0}$. Neutral impurity density is chosen to be $N=24\times p$, where $p$ is free hole concentration in GaN. This choice is not ad hoc;  measurements \cite{McLaurinJAP06} show that the neutral impurity density $N=N_{A}-p \approx 20-25 \times p$ for a wide range of  acceptor densities $N_{A}$.\\
\par
\begin{figure}[t]%
\includegraphics*[width=92 mm]{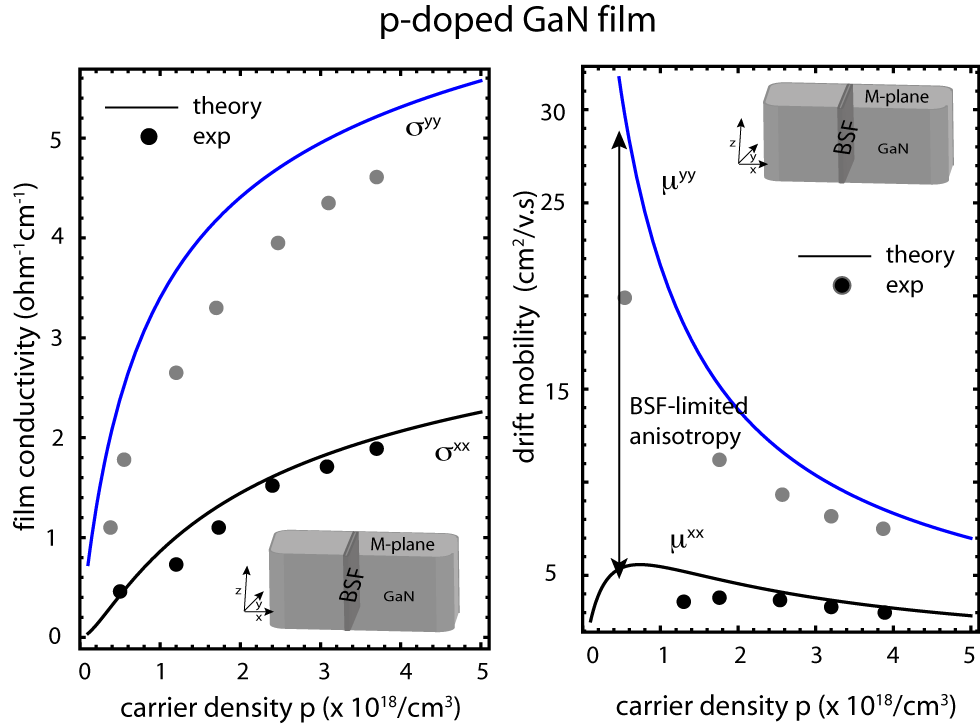}
\caption{%
  a) hole conductivity as a function of hole density: anisotropy is due to BSF b) hole mobility as a function of hole density. solid lines are theoretical values and solid circles are experimental values  from ref. 7.}
\label{Fig2}
\end{figure}

 Figure \ref{Fig2}a) shows the calculated (solid lines) variation of $\sigma_{xx}$ (along the $c$-axis) and $\sigma_{yy}$ (parallel to $a$ axis) as a function of hole density at room temperature juxtaposed with experimental \cite{McLaurinJAP06} data (filled circles). It is apparent from the figure that the presence of BSF causes an appreciable reduction of $\sigma_{xx}$ compared to $\sigma_{yy}$, resulting in anisotropic hole transport . The corresponding hole mobility $\mu_{ii}=\sigma_{ii}/pe$; $i=x,y$, is shown in Fig.\ref{Fig2}b). The anisotropy in hole mobility increases with decreasing hole density. This stems out from the fact that, in the limit $p\rightarrow 0$, the depletion width increases ($x_{d}\sim N_{A}^{-1/2}$) and one approaches the critical acceptor density limit $N_{A}^{cr}$. As $x_{d}$ increases, the strength of the barrier $S_{1}$ increases, and $T_{tr}(\epsilon)\rightarrow 0$. This results in vanishing $\mu_{xx}$ ($\sigma_{xx}$; see Fig.\ref{Fig2}).\\
 \par
 We next investigate the electron transport in n-doped samples in the presence of BSFs.  Experiment \cite{McLaurinJAP06} shows that electron mobility in m-plane GaN decreases with decreasing electron concentration. This phenomena is well described by charged dislocation scattering as pointed out by Weimann et. al \cite{WeimannJAP98}. The large lattice mismatch between GaN and the substrate (for GaN on SiC, lattice mismatch $\sim 2.9$\%) leads to high dislocation densities.  For m-plane GaN grown on foreign substrates, TEM images reveal \cite{LoAPL08,ZakharovPRB05} edge dislocation lines perpendicular to the m-plane ( along the [1$\bar{1}$00] direction) with non zero Burger's vector along [0001] direction.  Each dislocation line acts as an acceptor-like trap with a line charge density $\rho_{L}=e/a$, where $a$ is the $a$-lattice constant of GaN. The electrons moving in the perpendicular plane to the dislocation line effectively feel a Coulomb potential $V(r)=(\rho_{L}/2\pi\epsilon_{s})K_{0}(r/L_{D})$, where $K_{0}(\cdot\cdot\cdot)$ is the zeroth order Bessel function of 2nd kind and $L_{D}=\sqrt{k_{B}T\epsilon_{s}/e^{2}n^{\prime}}$ is the Debye screening length. Both free electrons, and bound electrons are taken account in the effective screening concentration ($n^{\prime}$) using Brook's \cite{BookSeeger} formula $n^{\prime}=2n-n^{2} /N_{D}$; where $n$ is the free carrier concentration. The momentum relaxation time ($\tau_{dis}$) due to dislocation lines has been calculated by several authors for three \cite{PodorPSS66,WeimannJAP98,LookPRL98} and two-dimensional electron gas \cite{DJAPL00} in GaN. In the presence of BSFs, the $x$ component of conductivity at room temperature due to dislocation scattering can be calculated as
\begin{equation}
\sigma_{xx}^{dis}=\Big(\frac{\epsilon_{s}}{e^{2}}\Big)^{2}\frac{3ne^{2}a^{2}(k_{B}T)^{3/2}}{4\pi\sqrt{m^{\star }} n_{dis}L_{D}}{\mathcal{I}}(n,\Delta E_{c}),
\end{equation}
\begin{figure}[t]%
\includegraphics*[width=94 mm]{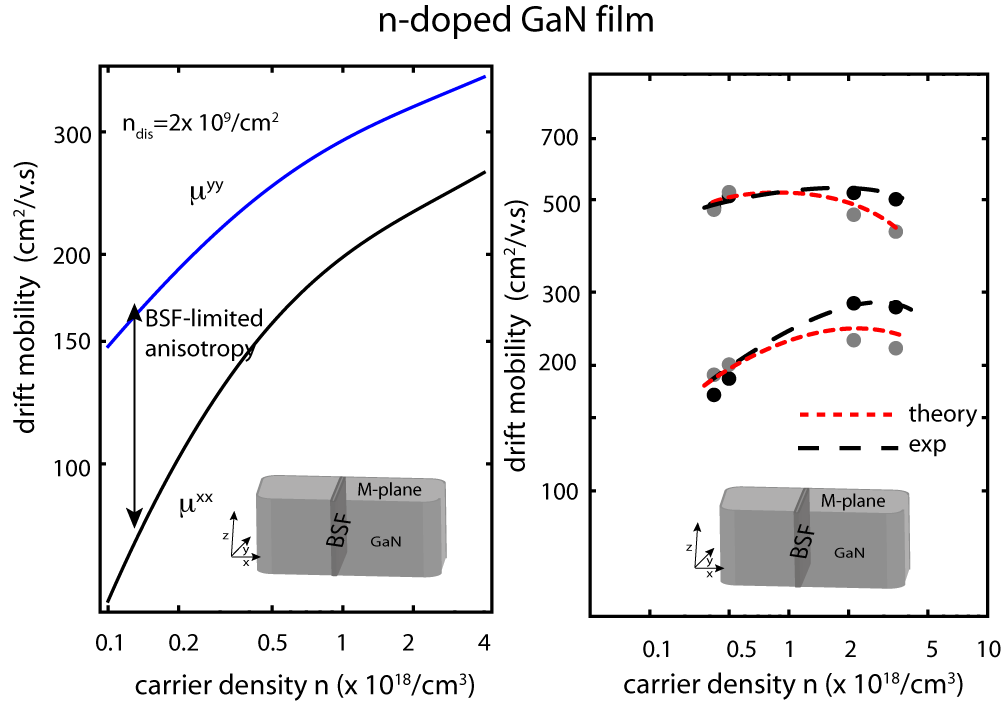}
\caption{%
  a) electron mobility as a function of electron density , b) comparison between theoretical values (dotted line) and experimental values (dashed line) of electron mobility at different values of electron concentrations. }
\label{Fig3}
\end{figure}
where, $n_{dis}$ is the dislocation density and ${\mathcal{I}}(n,\Delta E_{c})$ is a dimensionless integral. The complex nature of the of the integral ${\mathcal{I}}(n,\Delta E_{c})$ prevents an analytical evaluation and one must rely on numerical techniques.  For $\Delta E_{c}= 0.27$ eV \cite{StampflPRB98}, and in the non-degenerate limit,  ${\mathcal{I}}(n,0.27)\simeq 6 $, and $\mu_{xx}\sim \sqrt{n}$.  Fig. \ref{Fig3}a) shows the room temperature drift mobility with electron density for dislocation density $n_{dis}=2\times 10^{9}$/cm$^{2}$. It not only captures the mobility anisotropy due to BSF, but also the mobility variation with carrier concentration; the behavior is similar to those measured experimentally \cite{McLaurinJAP06,WeimannJAP98}. For calculations, we used $m_{xx}^{\star}=m_{yy}^{\star}=0.22 m_{0}$, and a resultant momentum relaxation time $\tau^{-1}=\tau_{dis}^{-1}+\tau_{NI}^{-1}+\tau_{imp}^{-1}+\tau_{ph}^{-1}$ is considered. Unlike p-doped GaN, the  unintentionally doped $C$ atoms serve as neutral impurities with a  measured density $N\sim 5\times 10^{16}$/cm$^{2}$ \cite{McLaurinJAP06}. Measurements of ref. 7 show free electron concentration using Hall technique is lower than actual dopant concentration in n-doped non-polar GaN samples. The difference between the dopant concentration and the free electron concentration is attributed to the acceptor traps associated with the dislocation lines. Moreover, the difference $(N_{D}-n)\sim n_{dis}$ varies from sample to sample indicating sample dependent density of dislocations.  Assuming a single trap per $a$-lattice constant, the volume density of trapped charge is $n_{dis}/a$; and a charge balance equation can be written as
\begin{equation}
n+\frac{n_{dis}}{a}=\frac{N_{D}}{1+\frac{n}{2N_{C}}\exp(E_{D}/k_{B}T)},
\label{Eq6}
\end{equation}
where, $N_{C}=2(m^{\star}k_{B}T/2\pi\hbar^{2})^{3/2}$ is the effective density of states and $E_{D}$ is the activation energy of the donor level.  The screening effect on the donor activation energy is taken account by the empirical formula $E_{D}=E_{D0}-\alpha N_{D}^{1/3}$ \cite{LookSSC97}, where the screening parameter $\alpha= 2.1\times 10^{-5}$ meV.cm and $E_{D0}=28$ meV \cite{LookSSC97}. Using Eq. \ref{Eq6} and values of $N_{D}$ and $n$ from ref. 7 calculated dislocation densities are found to be in the range of n$_{dis}\sim$ 1-10 x $10^{8}$ /cm$^{2}$. Electron mobilities is determined using the calculated values of sample-dependent dislocation densities and compared to experimentally measured values as shown in Fig. \ref{Fig3}b). At low carrier densities, theoretical values are in agreement with experimental data; but they differ at high carrier concentrations. \\
\par
An important observation is that the anisotropy in electron and hole mobilities decreases with increasing carrier concentration. At higher carrier densities, transmission of carriers occupying higher energy states approaches unity ($T_{tr}(\epsilon)\rightarrow 1$) leading to $\mu_{xx}\sim \mu_{yy}$.  
The corresponding carrier density is $n_{0}(p_{0})\simeq 1/3\pi^{2}(m^{\star}\Delta E_{c(v)}d/\hbar^{2})^{3}$ ($\sim 10^{19}$ /cm$^{3}$), above which BSF-induced transport anisotropy vanishes in non-polar GaN. At high acceptor concentrations, the energy levels of impurity atoms overlaps to form impurity bands. While conduction through this impurity band is important at low temperatures, at room temperature impurity band does not form an efficient transport path \cite{LancefieldJP01} and its exclusion from our model is quite justified. Also, the comparison between theoretically calculated drift mobility and experimentally measured Hall mobility is meaningful as at room temperature, the Hall factor $r_{H}\sim 1$ \cite{LookSSC97}. Our work presented here can be improved in three ways;  i) at low carrier densities $x_{d}$ is large; assumption of the delta function barrier breaks down and one should solve the Schroedinger equation numerically to obtain exact transmission coefficient $T_{tr1}(\epsilon)$, and ii) at low doping levels (when $x_{d}$ is large) or for high stacking fault density, carriers are confined in a quasi-two dimensional space rather than moving in three dimensions. Hence, one should use a two-dimensional analogue of Eq.\ref{eq4} for an exact evaluation of $\mu_{yy}$, and iii) by incorporating the contributions of quasi-localized \cite{RebanePSS97} electrons in the shallow state of the thin QW located in the fault region. \\
\par
High-quality (low defect densities) non-polar GaN is currently under development. It is apparent that with decreasing SF density, transport anisotropy will decrease and eventually vanishes for BSF free non-polar GaN films.  The theory presented here is not only applicable to GaN, but to any semiconductor (including other members of the nitride family such as InN \cite{HsiaoJAP10} and AlN \cite{UenoJJAP10}) having BSFs with proper choice of material parameters. For example, the semiconductor zinc-oxide (ZnO) shows SFs similar to GaN with $\sigma_{\pi}= .057$ C/m$^{2}$ and the band offsets \cite{YanPRB04} are $\Delta E_{c}=0.147$ eV  and $\Delta E_{v}=0.037$ eV.\\
\par
In summary, we have presented a transport theory in semiconductors containing a large number of basal stacking faults. The theory is applied to understand the experimentally observed transport anisotropy in m-plane GaN and a reasonable agreement between theory and experiment is obtained. Two critical limits of carrier concentration are derived, one where transport across a basal stacking fault ceases and another, where the presence of basal stacking faults can be ignored in the context of charge transport.\\
The authors would like to acknowledge  J. Verma, S. Ganguly (University of Notre Dame) for helpful discussions and National Science Foundation (NSF), Midwest Institute for Nanoelectronics Discovery (MIND) for the financial support for this work.

\end{document}